\documentclass[12pt]{article}

\usepackage[utf8]{inputenc}
\usepackage[T1]{fontenc}
\usepackage{mathptmx} 
\usepackage{setspace}
\usepackage[margin=2.5cm]{geometry}

\usepackage{graphicx}
\usepackage{float}
\usepackage{adjustbox}
\usepackage{tabularx}
\usepackage{makecell}
\usepackage{amsthm}
\usepackage[table]{xcolor} 
\usepackage{mathtools}     
\usepackage{amssymb}
\usepackage{bm}
\usepackage{textgreek}
\usepackage{booktabs}
\usepackage{subcaption}
\usepackage{ragged2e}
\usepackage{placeins}
\newcolumntype{Y}{>{\raggedright\arraybackslash}X}

\definecolor{capcol}{HTML}{0B7285}   
\definecolor{compcol}{HTML}{B03A2E}  
\definecolor{r2col}{HTML}{1E8449}    
\definecolor{capnum}{HTML}{12A8B3}   
\definecolor{compnum}{HTML}{E67E73}  
\definecolor{r2num}{HTML}{7DCEA0}    
\definecolor{softteal}{HTML}{E8F6F7}
\definecolor{softmint}{HTML}{EAF7EE}
\definecolor{softsalmon}{HTML}{FDECEA}
\definecolor{softblue}{HTML}{E6F2FF}
\definecolor{softgold}{HTML}{FFF7E6}
\definecolor{softviolet}{HTML}{F3EDFF}

\usepackage[numbers,sort&compress]{natbib}
\usepackage{hyperref}
\hypersetup{
    colorlinks=true,
    linkcolor=blue,
    citecolor=blue,
    urlcolor=blue
}

\title{\bfseries Variational Percolation Bounds for \emph{Cellular Membrane} Occlusion}
\author{
   Cesar Mello$^{1,*}$ \\
   Fernando Medina da Cunha, MD$^{2}$ \\
   \\
   $^1$Cosmo Physics Organization, São Paulo, Brazil \\
   $^2$Centro de Oncologia Campinas, São Paulo, Brazil \\
   *\texttt{cesar.mello@cosmophys.org}
}
\date{}

\begin{document}
\maketitle
\begingroup\sloppy

\begin{abstract}
Malignant membranes cluster nutrient transporters within glycan rich domains, sustaining metabolism through redundant intake routes. A theoretical framework links interfacial chemistry to transport suppression and energetic or redox collapse. The model unites a screened Poisson–Nernst–Planck electrodiffusion problem, an interfacial potential of mean force, and a reduced energetic redox module connecting flux to ATP/NADPH balance. From this structure, capacitary spectral bounds relate total flux to the inverse principal eigenvalue,
\[
J_{\mathrm{tot}}\!\le\!C\,e^{-\beta\chi_{\mathrm{eff}}}\mathcal{P}(\theta).
\]
Two near orthogonal levers geometry and field strength govern a linear suppression regime below a percolation-type knee, beyond which conductance collapses. A composite intake index,
\[
\Xi=w_{\mathrm{G}}J_{\mathrm{GLUT}}+w_{\mathrm{A}}J_{\mathrm{LAT/ASCT}}+w_{\mathrm{L}}J_{\mathrm{MCT}},
\]
dictates energetic trajectories: once below a maintenance threshold, ATP and NADPH fall jointly and redox imbalance drives irreversible commitment. Normal membranes, with fewer transport mouths and weaker fields, remain above this threshold, defining a natural selectivity window. The framework demonstrates existence, regularity, and spectral monotonicity for the self-adjoint PNP operator, establishing a geometric–spectral transition that links molecular parameters such as branching and sulfonation to measurable macroscopic outcomes with predictive precision.
\end{abstract}
\newcommand{\eqlarge}[1]{\begingroup\fontsize{14pt}{18pt}\selectfont #1\endgroup}

\section*{Introduction}
\begingroup\sloppy
\paragraph*{}
Solid tumors exhibit clustered nutrient transporters (GLUT1/3, LAT1/ASCT2, MCT1/4) embedded in glycan–rich microdomains \cite{23,34,8}. Inspired by this organization, a purely theoretical, membrane–centric two–stage \emph{lockdown} is formulated. In \emph{Stage~1}, a branched sulfonated PEG scaffold (Cap–PEG–SO$_3$) anchored by DSPE–PEG or chol–PEG imposes multichannel occlusion at transporter vestibules; in \emph{Stage~2}, an optional prodrug layer releases payloads under tumor–specific triggers (pH, GSH, protease) \cite{41,42,5}. The framework aims to map controllable interfacial chemistry onto transport and energetic/redox behavior in malignant versus non–malignant membranes.

\paragraph*{}
Despite extensive work on metabolic redundancy, no quantitative scheme links interfacial architecture and sulfonation to global transport suppression and energetic failure while remaining falsifiable. Classical single–route inhibition fails under redundant intake; a unified membrane–level theory is needed to describe \emph{multichannel} occlusion, identify the onset of conductance collapse, and define selectivity as operation below a percolation–type knee with moderated fields \cite{50,15}.

\paragraph*{}
The framework couples three layers:  
(i) a screened \textbf{Poisson–Nernst–Planck (PNP)} electrodiffusion problem on heterogeneous membranes with mixed boundaries, formulated in weak form under explicit regularity and screening hypotheses;  
(ii) an interfacial potential of mean force combining steric, electrostatic, hydration, dispersion, and multivalency terms;  
(iii) a reduced energetic–redox network mapping flux suppression onto ATP/NADPH balance and metabolic stability \cite{54,55,49}.  
Together, they connect tunable chemical knobs (branching, grafting, sulfonation) to spectral conductance and redox outcomes through variational inequalities.

\paragraph*{}
Two lightly orthogonal variables—geometric coverage $\theta$ and interfacial field $\chi_{\mathrm{eff}}$—govern the capacitary–spectral law
\[
\eqlarge{
J_{\mathrm{tot}}\;\le\; C\,e^{-\beta\chi_{\mathrm{eff}}}\,P(\theta),
}
\]
with a percolation knee at $\theta=\eta_c$. Variational–spectral arguments describe how occlusion and field amplification raise the principal eigenvalue and reduce conductance.  
Selectivity follows from remaining below $\eta_c$ in normal contexts, while malignant membranes exceed it.  
The theory yields falsifiable slopes linking molecular design to measurable observables (fluxes, ATP, NADPH, ECAR, caspases) without invoking experimental claims.  
The next subsection, \emph{Controlled Orthogonality: Geometry vs.\ Interfacial Field}, formalizes why $\theta$ and $\chi_{\mathrm{eff}}$ act as quasi–independent levers defining a predictable and safe design space.
\endgroup
\begingroup\sloppy
\section*{Theoretical Foundations}

\subsection*{Scope and Modeling Layers}
A membrane–centric mechanism is formalized in three coupled layers:
(i) an \emph{electro-diffusive} boundary layer describing metabolite approach to transporter vestibules under screened fields;
(ii) an \emph{interfacial interaction} layer combining steric, electrostatic, hydrogen-bonding, dispersion, and hydration/depletion forces between the CAP scaffold and malignant microdomains;
(iii) a \emph{bioenergetic/network} layer where suppressed transport translates into ATP/NADPH decline and contraction of feasible flux states. The objective is a closed physics–chemistry map from molecular design variables to macroscopic flux bounds and dynamical outcomes \cite{54,55,34}.

\subsection*{Geometry, Notation, and Assumptions}
A membrane patch $\Omega\subset\mathbb{R}^2$ hosts clustered glycan–transporter microdomains. Extracellular mouths of $\mathcal{T}=\{\mathrm{GLUT1/3},\mathrm{LAT1},\mathrm{ASCT2},\mathrm{MCT1/4}\}$ are modeled as absorbing windows $\{\Gamma_j\}_{j=1}^N$ at centers $\{x_j\}$. CAP-occluded mouths are replaced by Dirichlet obstacles $\{\Gamma^{\rm cap}_j\}$ with effective radius $R_{\rm eff}$. Mouth centers follow a clustered Poisson process of intensity $\rho$ and cluster radius $r_c$, with parameters taken in physiological ranges unless noted \cite{23,34}.

\subsection*{Electro-diffusive Layer: Poisson–Nernst–Planck with Screening}
For each species $k$ (glucose, leucine, glutamine, lactate, protons), concentrations $c_k(x,t)$ and potential $\phi(x,t)$ satisfy
{\begingroup\fontsize{14pt}{18pt}\selectfont
\begin{align}
\partial_t c_k &= \nabla\!\cdot\!\Big(D_k \nabla c_k + \mu_k z_k c_k \nabla \phi\Big), \label{eq:PNP1}\\
-\,\nabla\!\cdot(\varepsilon \nabla \phi) &= \sum_k z_k c_k - \rho_f, \label{eq:PNP2}
\end{align}
\endgroup}
with $D_k$ diffusivity, $\mu_k$ mobility, $z_k$ valence, $\varepsilon$ permittivity, and $\rho_f$ fixed charge. Absorbing mouths impose flux on $\Gamma_{\rm abs}$, CAP-occluded mouths enforce $c_k{=}0$ and $\phi{=}\psi_s$ on $\Gamma^{\rm cap}_j$, and no-flux Neumann conditions hold on $\Gamma_{\rm N}$ \cite{54,55,49}.

\paragraph{Functional setting and well-posedness.}
Let $\partial\Omega=\Gamma_{\rm abs}\cup\Gamma_{\rm cap}\cup\Gamma_{\rm N}$ be a mixed boundary decomposition with $\Omega$ bounded and $C^{1,\alpha}$. Under Debye–Hückel screening ($|\phi|/\phi_T\ll1$) and the Einstein relation $\mu_k=D_k/(k_BT)$, the linearized steady operator on $H^1_{\Gamma_{\rm D}}(\Omega)$ (with $\Gamma_{\rm D}=\Gamma_{\rm abs}\cup\Gamma_{\rm cap}$) is symmetric, coercive, and has compact resolvent, yielding a unique weak solution (steady) and an analytic semigroup (transient). This ensures that fluxes and access factors below are well defined within the mixed-boundary PNP framework.

\paragraph{Scaling.}
With $L\!\sim\!10$–100\,nm, $\phi_T=k_BT/e$, and nondimensional variables $\tilde{x}=x/L$, $\tilde{\phi}=\phi/\phi_T$:
\[
\eqlarge{
\delta_D=\lambda_D/L,\quad
\chi_k=z_k\psi_s/\phi_T,\quad
\mathrm{Pe}_k=\tfrac{\|\mu_k z_k c_k \nabla \phi\|L}{\|D_k \nabla c_k\|}.
}
\]
Screening is effective for $\delta_D\!\lesssim\!1$; strong anionic gating corresponds to $\chi_k\!\gg\!1$ \cite{54}.
\paragraph{Boundary fluxes and linearized screening.}
The normal flux for species $k$ is $J_k^n=\big(-D_k \nabla c_k - \mu_k z_k c_k \nabla \phi\big)\!\cdot n$. On $\Gamma_{\rm abs}$ the (active) absorbing/uptake condition is imposed through $J_k^n$, on $\Gamma_{\rm cap}$ the constraints $c_k=0$ and $\phi=\psi_s$ are enforced, and on $\Gamma_{\rm N}$ the no-flux condition $J_k^n=0$ holds. In the Debye–Hückel regime, linearizing \eqref{eq:PNP2} about $(c_{k,\infty},0)$ yields the screened Poisson operator
\[
\eqlarge{
-\,\varepsilon\Delta \phi + \varepsilon\kappa_D^2\,\phi \;=\; \sum_k z_k\,(c_k-c_{k,\infty}),\qquad \kappa_D=\lambda_D^{-1},
}
\]
so that $\phi$ decays on the Debye scale and couples weakly to far-field concentrations when $\delta_D\lesssim 1$.

\paragraph{Weak formulation (linearized steady state).}
Let $V=H^1_{\Gamma_{\rm D}}(\Omega)$ with $\Gamma_{\rm D}=\Gamma_{\rm abs}\cup\Gamma_{\rm cap}$. For fixed $\phi$, the drift–diffusion operator admits the bilinear form
\[
a_k(u,v)=\int_\Omega D_k \nabla u\!\cdot\!\nabla v\,dV \;+\; \int_\Omega \mu_k z_k \,\bar c_k\,\nabla\phi\!\cdot\!\nabla v\,dV,
\]
with $\bar c_k$ a positive reference profile. Coercivity of the symmetric part and compact embedding imply existence/uniqueness by Lax–Milgram/Fredholm; maximum principles control $c_k\ge 0$ under absorbing Dirichlet/flux conditions.

\paragraph{Narrow-access asymptotics and capacity.}
When $\Gamma_{\rm abs}$ consists of small windows, the total steady flux scales with the boundary capacity,
\[
\eqlarge{
J_k \;\sim\; \kappa_k^{0}\,\mathrm{Cap}(\Gamma_{\rm abs})\,c_{k,\infty}\,e^{-\beta \chi_k},
}
\]
recovering the factorization used below: steric occlusion shrinks the effective accessible set (geometry, via $\theta$), while screening and surface potential modulate entrance through a Boltzmann factor (field, via $\chi_{\rm eff}$). This matches the spectral picture $J_{\rm tot}\asymp \lambda_1^{-1}$ and underlies the linear–tunable regime away from the percolation knee.

\begin{figure}[H]
    \centering
    \includegraphics[width=0.9999\linewidth]{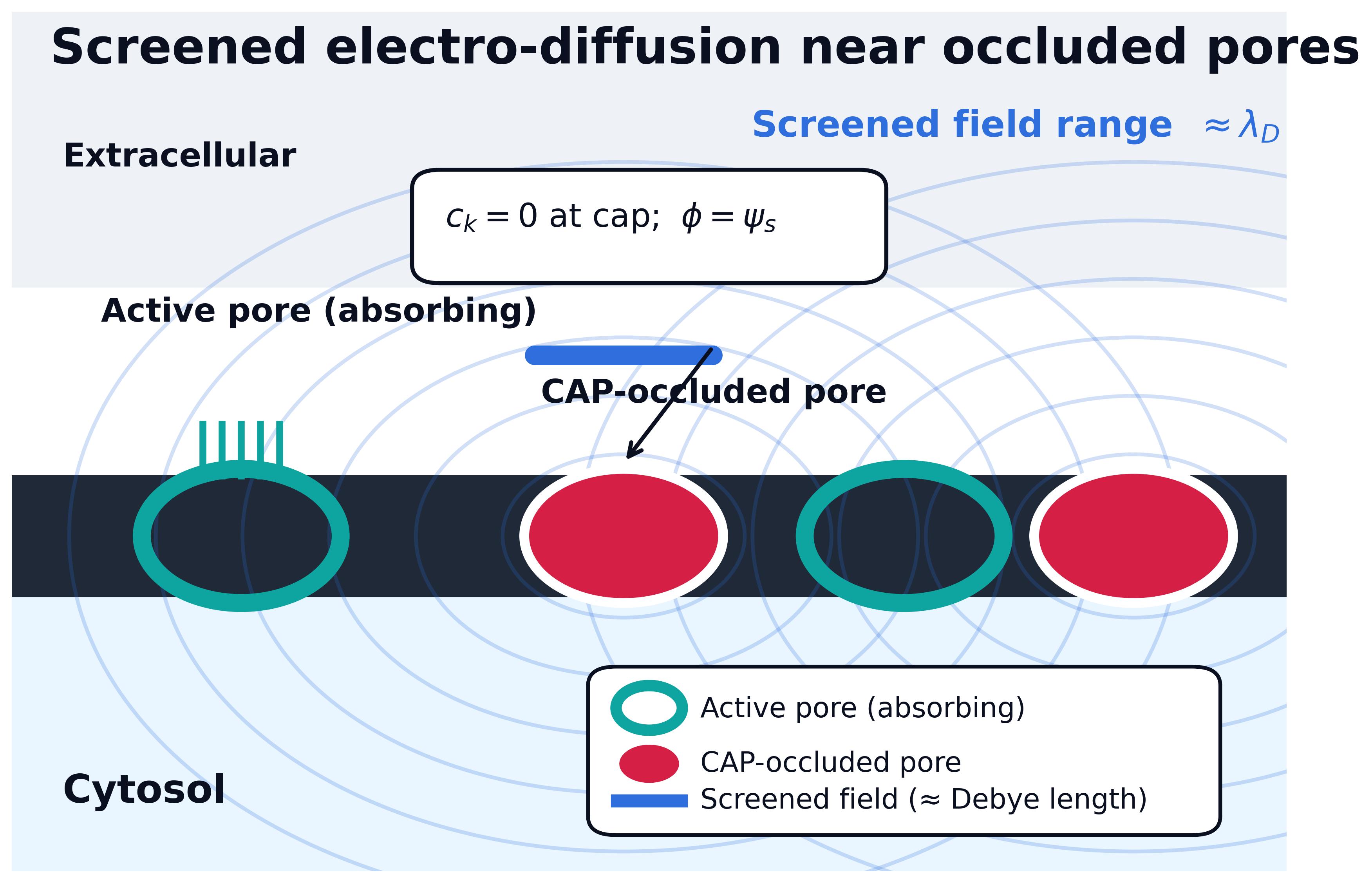}
    \caption{\textbf{Screened electro-diffusion near occluded pores.}
    A membrane band with alternating absorbing mouths and CAP-occluded mouths ($c_k{=}0$, $\phi{=}\psi_s$). The interfacial field decays over $\lambda_D$.}
    \label{fig:PNP_Schematic}
\end{figure}

Far-field transport is set by screened PNP, while near-field access is shaped by absorbing versus occluded mouths. Within the halo of an occluded pore, approach is modulated by a potential of mean force. The resulting access factor $\Pi_k=\exp[-\beta U_k]$, combined with geometric coverage $\theta$, yields
\[
\eqlarge{
J_k \;\approx\; \kappa_k^{0}\,(1-\theta)\,\Pi_k\,A_{\mathrm{act}}\,c_{k,\infty},
}
\]
with $\kappa_k^{0}$ intrinsic access coefficient and $A_{\mathrm{act}}$ active absorbing area. Steric geometry ($\theta$) acts broadly, charge ($\psi_s,\lambda_D$) selectively \cite{54,53,50}.

\subsection*{Interfacial Interaction Manifold}
Upon tumor-gated activation, CAP experiences a composite potential $U(r,\Omega)$ including:
steric exclusion by PEG brushes; screened electrostatics from sulfonates ($\psi_0$); hydrogen bonding and dispersion; hydration/depletion restructuring; and multivalency reducing $k_{\rm off}$. These align with bioconjugation thermodynamics and glycan-targeting chemistries \cite{50,53,57}.
\begin{figure}[H]
    \centering
    \includegraphics[width=0.92\linewidth]{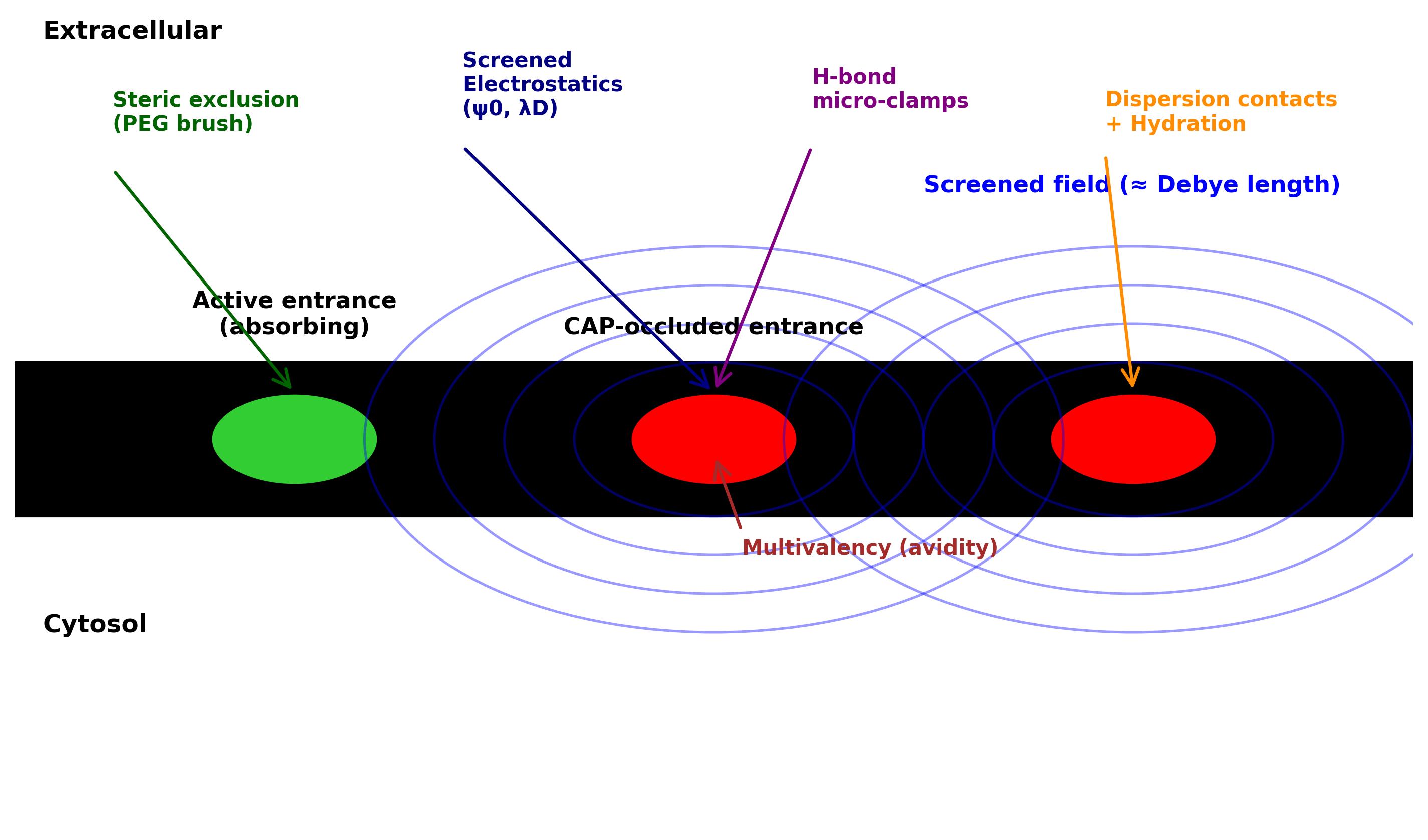}
    \caption{\textbf{Interfacial interaction manifold at a glycan–transporter cluster.}
    A branched PEG–SO$_3^-$ cap generates a composite potential-of-mean-force, stabilizing residence and narrowing access.}
    \label{fig:cap_interface_energy}
\end{figure}

Two orthogonal levers emerge: geometry (coverage $\theta$ via brush height/density) and field strength $\chi_{\mathrm{eff}}\!\propto|\psi_0|$ (via sulfonation/ionic strength). Near an entrance,
\[
\eqlarge{
U_k=U_{\text{steric}}+U_{\text{el}}+U_{\text{HB}}+U_{\text{disp}}+U_{\text{hydr}}-n_{\text{val}}\Delta G_{\text{avid}},
\quad \Pi_k=e^{-\beta U_k},
}
\]
so that
\[
\eqlarge{
\kappa_k^{\mathrm{eff}}=\kappa_k^{0}(1-\theta)\Pi_k,\qquad
J_k \approx \kappa_k^{\mathrm{eff}}\,A_{\mathrm{act}}\,c_{k,\infty}.
}
\]
This approximate orthogonality mirrors classical handles in surface engineering \cite{50,58,53}.

\subsection*{Chemistry$\to$Boundary Map}
Brush height scales as
\[
\eqlarge{
h_{\rm brush}\sim aN_{\rm seg}(\sigma_{\rm PEG}a^2)^{1/3}g(N_{\rm br}),\qquad
R_{\rm eff}=R_0+h_{\rm brush}+\alpha_{\rm solv} I^{-1/2}.
}
\]
Sulfonation and pH set charge and potential,
\[
\eqlarge{
\sigma_s=-\,\tfrac{eN_{\rm SO_3}\alpha_{\rm diss}(pH)}{A},\quad
\psi_0=\psi_s+\sigma_s/C_S,\quad
\psi_s=2\phi_T\,\mathrm{asinh}\!\Big(\tfrac{\sigma_s}{\sqrt{8\varepsilon k_BT c_\infty}}\Big).
}
\]
Hence $\theta$ (geometry) and $\psi_0$ (field) offer near-orthogonal control \cite{53,50,58}.

\subsection*{Variational and Spectral Bounds}
The PNP free-energy functional
\[
\eqlarge{
\mathcal{F}[\{c_k\},\phi]
= \sum_k \!\int_\Omega \!\Big(D_k |\nabla \sqrt{c_k}|^2 + z_k c_k \phi \Big)\,dV
+ \tfrac{\varepsilon}{2}\!\int_\Omega |\nabla \phi|^2\,dV
}
\]
minimized under mixed boundaries yields the capacitary–variational inequality
\[
\eqlarge{
J_{\rm tot}\le C(\{D_k\},\Omega)\,e^{-\beta\chi_{\rm eff}}\mathcal{P}(\theta).
}
\]
For the linearized operator $L_\chi=-\nabla\!\cdot(D\nabla)+\sigma(\chi_{\rm eff})$ on $H^1_{\Gamma_{\rm D}}(\Omega)$, let $\lambda_1(\theta,\chi_{\rm eff})$ be the principal Dirichlet eigenvalue. By domain monotonicity (increasing Dirichlet set as $\theta\uparrow$) and screening ($\sigma$ increasing with $\chi_{\rm eff}$), $\lambda_1$ increases with $\theta$ and $\chi_{\rm eff}$; consequently,
\[
\eqlarge{
J_{\rm tot}\ \asymp\ \frac{\mathrm{Cap}(\Gamma_{\rm abs})}{\lambda_1(\theta,\chi_{\rm eff})}
\ \lesssim\ C\,e^{-\beta\chi_{\rm eff}}\,\mathcal{P}(\theta),
}
\]
so that the effective spectral conductance $1/\lambda_1$ shrinks with occlusion and interfacial field, in agreement with the variational law \cite{54,55,49}.

\subsection*{Parallel with Percolation}
The geometric term scales as
\[
\eqlarge{
\mathcal{P}(\theta)\sim(1-\eta_c)\Big(\tfrac{1-\theta}{1-\eta_c}\Big)^t,\quad t\simeq1.3,
}
\]
matching 2D conductivity exponents. Together,
\[
\eqlarge{
J_k\sim J_k^0(1-\theta)e^{-\beta\chi_{\rm eff}}
\Big(\tfrac{1-\theta}{1-\eta_c}\Big)^{tH(\theta-\eta_c)},
}
\]
with $H$ the Heaviside step \cite{54,55}.

\subsection*{Bioenergetic–Redox Coupling}
Let $A(t),P(t)$ denote ATP and NADPH. A reduced model,
{\begingroup\fontsize{14pt}{18pt}\selectfont
\begin{align}
\dot A &= y_{\rm gly}J_{\rm GLUT}+y_{\rm ox}J_{\rm TCA}-k_{\rm ATP}\tfrac{A}{K_A+A}, \\
\dot P &= y_{\rm PPP}J_{\rm GLUT}-k_{\rm scav}\tfrac{R_{\rm ROS}}{K_P+P},
\end{align}
\endgroup}
receives fluxes bounded by the variational law. Collapse occurs when $(A,P)$ cross a separatrix defined by simultaneous throttling of intake and lactate handling \cite{31,32,33}.

\subsection*{Constraint-Based Network and Identifiability}
Transport caps shrink the feasible polytope $Sv=0$ via upper bounds $\overline v_{\rm uptake,k}(J_k)$. Shadow prices reveal emergent bottlenecks (notably lactate export). Sobol indices rank $R_{\rm eff}$ and $\psi_0$ for identifiability, tying observables to chemical levers \cite{48,47}.

\subsection*{Design–Space Implications}
The mechanism addresses malignant membranes with clustered, redundant nutrient mouths. Orthogonal tuning of $\theta$ and $\chi_{\rm eff}$ affords quasi-linear control up to $\eta_c$, with abrupt collapse beyond. A theoretical selectivity window arises by operating below $\eta_c$ and moderating $\chi_{\rm eff}$ in normal contexts \cite{20,23,34}.

\subsection*{Cap–Prodrug Integration: A Two-Stage Apoptosis Switch}
An extension couples starvation with gated sabotage. Stage~1: occlusion fraction $f(t)$ follows
\[
\eqlarge{
\dot f = k_{\rm on}[{\rm CAP}](1-f)-k_{\rm off}f,\qquad
f^\star=\tfrac{k_{\rm on}[{\rm CAP}]}{k_{\rm on}[{\rm CAP}]+k_{\rm off}}.
}
\]
Stage~2: payload activation $g(t)$,
\[
\eqlarge{
\dot g=k_{\rm rel}(pH,[{\rm GSH}],[{\rm MMP}])\,f(t)(1-g),\quad
\Phi_{\rm tox}(t)=\int_0^t\gamma f(\tau)g(\tau)\,d\tau.
}
\]
Coupling to reserves,
{\begingroup\fontsize{14pt}{18pt}\selectfont
\begin{align}
\dot A &= y_{\rm gly}(1-f)J_{\rm GLUT}+y_{\rm aa}(1-f)J_{\rm LAT/ASCT}-\alpha_A A, \\
\dot P &= y_{\rm PPP}(1-f)J_{\rm GLUT}-\alpha_P P-\beta\Phi_{\rm tox}(t),
\end{align}
\endgroup}
drives irreversibility once $(A,P)$ cross $\Sigma=\{A<A_{\rm crit},P<P_{\rm crit}\}$. A Lyapunov functional
\[
\eqlarge{
\mathcal{L}[f,g,A,P]=\tfrac{1}{2}(A^2+P^2)+\lambda_1 f^2+\lambda_2 g^2
}
\]
decreases monotonically under joint engagement, producing hysteresis \cite{41,42,44}.

\medskip
\noindent\textbf{Safety window.}
Normal contexts require
\[
\eqlarge{
J_{\rm GLUT}^{\rm basal}(1-f)\ge \eta_{\rm safe}J_{\rm GLUT}^{\rm basal},\qquad
\Phi_{\rm tox}^{\rm normal}(t)\le \phi_{\rm safe},
}
\]
with $f<\eta_c$ and subthreshold $g(t)$ \cite{31,33}.

\begin{table}[H]
\centering
\caption{\textbf{Stage~1 ligand–epitope interaction: illustrative design targets used in modeling (not experimental measurements).}}
\label{tab:ligand_params}
\rowcolors{2}{softblue}{white}
\begin{tabularx}{0.95\linewidth}{@{} l c c c c @{}}
\toprule
\rowcolor{softteal}
\textbf{Parameter} & \textbf{Symbol} & \textbf{Malignant} & \textbf{Non-malignant} & \textbf{$\Delta$ (Malig.–Non-malig.)} \\
\midrule
$K_D$ & & \cellcolor{softsalmon} 1--5\,nM & \cellcolor{softmint} $>$500\,nM & $\lesssim -495$\,nM \\
$k_{\text{on}}$ & & \cellcolor{softsalmon} $1.0\times10^6$ M$^{-1}$s$^{-1}$ & \cellcolor{softmint} $<5\times10^4$ M$^{-1}$s$^{-1}$ & $>\,9.5\times10^5$ M$^{-1}$s$^{-1}$ \\
$k_{\text{off}}$ & & \cellcolor{softsalmon} $<5\times10^{-4}$ s$^{-1}$ & \cellcolor{softmint} $>2.5\times10^{-2}$ s$^{-1}$ & $<\,-2.45\times10^{-2}$ s$^{-1}$ \\
$\tau_{\text{res}}$ & & \cellcolor{softsalmon} $>30$ min & \cellcolor{softmint} $<40$ s & $>\,29.3$ min \\
$N_{\text{ep}}$ & & \cellcolor{softsalmon} $(2{-}5)\times10^5$ & \cellcolor{softmint} $<5\times10^3$ & $\gtrsim (1.95{-}4.95)\times10^5$ \\
$K_D^{\text{app}}$ & & \cellcolor{softsalmon} $<1$\,nM & \cellcolor{softmint} $>200$\,nM & $<\,-199$\,nM \\
\bottomrule
\end{tabularx}
\end{table}
\vspace{4pt}
\noindent\textit{} The last column reports a directional contrast
\[
\Delta(X)\;:=\;X_{\text{Malignant}}-X_{\text{Non-malignant}},
\]
written as a \emph{conservative bound} when entries are given as ranges or with inequality signs. For quantities where \emph{smaller} is better affinity or slower loss (e.g., $K_D$ and $k_{\text{off}}$), a \textbf{negative} $\Delta$ indicates tighter binding/slower dissociation in malignant contexts. For quantities where \emph{larger} is favorable (e.g., $k_{\text{on}}$, $\tau_{\text{res}}$, $N_{\text{ep}}$), a \textbf{positive} $\Delta$ indicates malignant advantage. Because several cells use $<$ or $>$, the numeric $\Delta$ values are bounds, not exact differences.

\begin{itemize}
\item \textbf{$K_D$ (equilibrium dissociation constant).} For a 1:1 site, $K_D=\dfrac{k_{\text{off}}}{k_{\text{on}}}$ (M). Smaller $K_D$ means higher affinity. Table values (1–5\,nM vs $>$500\,nM) imply at least~$\sim\!100\times$ tighter binding in malignant membranes. The reported $\Delta$ reflects a conservative difference (e.g., $\lesssim-495$\,nM); an affinity \emph{fold gain} is better expressed as
\[
\mathrm{AF}_{K_D}\;:=\;\frac{K_D^{\text{non}}}{K_D^{\text{malig}}}\;\gtrsim\;\frac{500}{5}=100.
\]
\item \textbf{$k_{\text{on}}$ (association rate).} Units M$^{-1}$s$^{-1}$. Larger $k_{\text{on}}$ accelerates capture. With $1.0\times10^6$ vs $<5\times10^4$, the conservative fold-change is
\[
\mathrm{FC}_{k_{\text{on}}}\;\gtrsim\;\frac{10^6}{5\times10^4}=20.
\]
\item \textbf{$k_{\text{off}}$ (dissociation rate).} Units s$^{-1}$. Smaller $k_{\text{off}}$ stabilizes residency. With $<5\times10^{-4}$ vs $>2.5\times10^{-2}$,
\[
\mathrm{FC}_{k_{\text{off}}}\;\lesssim\;\frac{5\times10^{-4}}{2.5\times10^{-2}}=0.02,
\]
i.e., malignant dissociation is at least~$50\times$ slower.
\item \textbf{$\tau_{\text{res}}$ (residence time).} For a single-exponential off-process, $\tau_{\text{res}}\approx 1/k_{\text{off}}$. Thus $>30$\,min vs $<40$\,s implies
\[
\mathrm{FC}_{\tau_{\text{res}}}\;\gtrsim\;\frac{30\,\mathrm{min}}{40\,\mathrm{s}}\approx 45.
\]
\item \textbf{$N_{\text{ep}}$ (epitope density per cell).} Higher $N_{\text{ep}}$ raises effective local rebinding and multivalency. With $(2{-}5)\times10^5$ vs $<5\times10^3$,
\[
\mathrm{FC}_{N_{\text{ep}}}\;\gtrsim\;\frac{2\times10^5}{5\times10^3}=40\quad(\text{and up to }100\times \text{ at the upper range}).
\]
\item \textbf{$K_D^{\text{app}}$ (apparent, multivalent/avidity).} Multivalency and rebinding reduce the apparent dissociation constant relative to the monovalent $K_D$ (chelate effect), effectively lowering $k_{\text{off}}$ to $k_{\text{off}}^{\text{eff}}$. With $<1$\,nM vs $>200$\,nM,
\[
\mathrm{AF}_{K_D^{\text{app}}}\;\gtrsim\;\frac{200}{1}=200.
\]
\end{itemize}

\noindent\textbf{Conventions and caveats.} (i) When a cell shows “$<$” or “$>$”, $\Delta$ is reported as a bound consistent with those signs; it \emph{cannot} be a precise arithmetic difference. (ii) Because these parameters span orders of magnitude, a log-change is often more interpretable:
\[
\Delta_{\log_{10}}(X)\;:=\;\log_{10}X_{\text{Malig.}}-\log_{10}X_{\text{Non-malig.}}.
\]
For affinity, one typically reports $\Delta_{\log_{10}}(K_D)$ with the sign flipped (so that larger is tighter). (iii) Units: $K_D$ and $K_D^{\text{app}}$ in nM; $k_{\text{on}}$ in M$^{-1}$s$^{-1}$; $k_{\text{off}}$ in s$^{-1}$; $\tau_{\text{res}}$ in seconds/minutes; $N_{\text{ep}}$ is dimensionless (count per cell).

\medskip
\noindent\textbf{Physiological reading.} Larger $N_{\text{ep}}$ and $k_{\text{on}}$, together with smaller $k_{\text{off}}$, prolong membrane residency (higher $\tau_{\text{res}}$) in malignant membranes. In the modeling map, this increases effective coverage $\theta$ (more persistent occupancy of vestibules) and, for sulfonated scaffolds, strengthens the interfacial field proxy $\chi_{\mathrm{eff}}$ via local charge density. Both effects reduce spectral conductance and push operation toward the percolation knee, explaining the stronger suppression predicted for malignant versus non-malignant contexts.

\subsection*{Biphasic Death Trajectory}
A biphasic sequence follows occlusion. \emph{Phase~I} (energetic/redox shortfall) precedes \emph{Phase~II} (mitochondrial commitment and execution). Let
\[
\Xi(t)=w_{\rm G}J_{\mathrm{GLUT}}(t)+w_{\rm A}J_{\mathrm{LAT/ASCT}}(t)+w_{\rm L}J_{\mathrm{MCT}}(t)
\]
be the composite intake index (cf.\ earlier definition), and define maintenance thresholds $A_{\rm maint},P_{\rm maint}>0$ for ATP and NADPH. A minimal reduced model is
\begin{align}
\dot A &= s_A\,\Xi(t)-\alpha_A A, \qquad A(0)=A_0, \label{eq:A_dyn}\\
\dot P &= s_P\,J_{\mathrm{GLUT}}(t)-\alpha_P P-\gamma R, \qquad P(0)=P_0, \label{eq:P_dyn}\\
\dot R &= r_0 + r_1(1-M) - \delta_PP - \delta_R R, \qquad R(0)=R_0, \label{eq:R_dyn}\\
\dot M &= -k_{\Delta\Psi}\,\big(\rho_R R - \rho_A A - \rho_P P\big)_+ , \qquad M(0)=M_0, \label{eq:M_dyn}\\
\dot C &= k_c\,H(M_{\rm th}-M)\,(1-C)-\delta_c C, \qquad C(0)=0, \label{eq:C_dyn}
\end{align}
where $A,P$ denote ATP and NADPH, $R$ a lumped ROS/oxidative stress variable, $M\in[0,1]$ the mitochondrial polarization (1 = polarized), $C\in[0,1]$ caspase activation, $H$ the Heaviside step, $(\cdot)_+=\max\{\cdot,0\}$, and all parameters are positive. Occlusion enters only through $\Xi$ and $J_{\mathrm{GLUT}}$, which obey the capacitary–spectral law and bounds above.

\paragraph*{Link to ligand table (malignant vs.\ non-malignant).}
Denote by $\Delta(\cdot)= (\cdot)_{\rm Malignant}-(\cdot)_{\rm Non}$ the directional contrast in Table~\ref{tab:ligand_params}. The kinetics deltas
\[
\Delta k_{\rm on}>0,\quad \Delta k_{\rm off}<0,\quad \Delta \tau_{\rm res}>0,\quad \Delta N_{\rm ep}>0,\quad \Delta K_D^{\rm app}<0
\]
imply higher steady occupancy of vestibules in malignant membranes. With a standard surface-binding surrogate,
\begin{equation}
\dot f = k_{\rm on}N_{\rm ep}[{\rm CAP}]_{\rm surf}(1-f)-k_{\rm off}f,\qquad 
f^\star=\frac{k_{\rm on}N_{\rm ep}[{\rm CAP}]_{\rm surf}}{k_{\rm on}N_{\rm ep}[{\rm CAP}]_{\rm surf}+k_{\rm off}}
=\frac{1}{1+\dfrac{K_D^{\rm app}}{[{\rm CAP}]_{\rm surf}}}, 
\label{eq:fstar}
\end{equation}
so that $\Delta f^\star>0$ for any fixed $[{\rm CAP}]_{\rm surf}>0$. Effective geometric coverage and interfacial field are then lifted by occupancy:
\begin{equation}
\theta(t)=\theta_{\rm base}+\theta_{\max} f(t),\qquad 
\chi_{\rm eff}(t)=\chi_{\rm base}+\alpha_s\,\sigma_s\,f(t), \label{eq:theta-chi-map}
\end{equation}
yielding $\Delta\theta\ge \theta_{\max}\Delta f^\star>0$ and $\Delta\chi_{\rm eff}\ge \alpha_s\sigma_s\Delta f^\star>0$ (for fixed sulfonation density $\sigma_s$). Through the central law
\[
J_k^{\rm eff}=J_k^{0}\,(1-\theta)\,e^{-\beta \chi_{\rm eff}},
\]
the malignant-to-nonmalignant flux ratio obeys
\begin{equation}
\frac{J_k^{\rm malig}}{J_k^{\rm non}}\;\le\;
\underbrace{\frac{1-\theta_{\rm non}-\Delta\theta}{1-\theta_{\rm non}}}_{\text{steric loss}}
\cdot 
\underbrace{\exp\!\big[-\beta\,\Delta\chi_{\rm eff}\big]}_{\text{field penalty}}
\;=:\;\mathcal{S}_k(\Delta\theta,\Delta\chi_{\rm eff})\;<\;1, \label{eq:Sk}
\end{equation}
hence $\Xi_{\rm malig}(t)\le \sum_k w_k\,\mathcal{S}_k\,J_k^{\rm non}(t)$. The inequalities are conservative when table entries carry “$<$/$>$”; they still imply $\Xi_{\rm malig}<\Xi_{\rm non}$ for the same $[{\rm CAP}]_{\rm surf}$.

\paragraph*{Phase I (energetic/redox).}
If $\Xi(t)\equiv \Xi$ is (piecewise) constant on a timescale $\tau_1\ll\alpha_A^{-1}$, \eqref{eq:A_dyn} has the closed form
\[
A(t)=A_\infty+(A_0-A_\infty)e^{-\alpha_A t},\qquad A_\infty=\frac{s_A}{\alpha_A}\,\Xi.
\]
The \emph{tipping condition} $A(t)\downarrow A_{\rm maint}$ is crossed at
\[
t_{A\to{\rm maint}}
=\frac{1}{\alpha_A}\log\!\left(\frac{A_0-A_\infty}{A_{\rm maint}-A_\infty}\right),
\quad\text{provided }A_\infty<A_{\rm maint}.
\]
Because $\Xi_{\rm malig}=\mathcal{S}_{\rm G}\,w_{\rm G}J_{\rm GLUT}^{\rm non}+\mathcal{S}_{\rm A}\,w_{\rm A}J_{\rm LAT/ASCT}^{\rm non}+\mathcal{S}_{\rm L}\,w_{\rm L}J_{\rm MCT}^{\rm non}$ with $\mathcal{S}_k<1$ from \eqref{eq:Sk}, $A_\infty^{\rm malig}<(s_A/\alpha_A)\,\Xi_{\rm non}$ and $t_{A\to{\rm maint}}$ shortens monotonically with $(\Delta\theta,\Delta\chi_{\rm eff})$.
Analogously, if $J_{\mathrm{GLUT}}(t)\equiv J_{\mathrm{GLUT}}$ on $\tau_1$, then
\[
P(t)=P_\infty+(P_0-P_\infty)e^{-\alpha_P t}-\gamma\!\int_0^t e^{-\alpha_P(t-\tau)}R(\tau)\,d\tau,\quad
P_\infty=\frac{s_P}{\alpha_P}\,J_{\mathrm{GLUT}},
\]
so any lower bound $R(\tau)\ge \underline R>0$ yields
\[
P(t)\le P_\infty+(P_0-P_\infty)e^{-\alpha_P t}-\frac{\gamma\,\underline R}{\alpha_P}\,\big(1-e^{-\alpha_P t}\big),
\]
expediting $P(t)\downarrow P_{\rm maint}$. The \emph{Phase~I} window ends once either
\[
J_{\mathrm{GLUT}}+J_{\mathrm{LAT/ASCT}}<J_{\min}\quad\text{and}\quad P(t)<P_{\mathrm{crit}},
\]
at which point compensatory fluxes cannot prevent mitochondrial engagement \cite{31,32,51}.

\paragraph*{Phase II (commitment and execution).}
Given \eqref{eq:R_dyn}–\eqref{eq:M_dyn}, if $A\le A_{\rm maint}$ and $P\le P_{\rm maint}$ on $[t_0,\infty)$ and $R(t)\ge \underline R>0$, then $\dot M\le -k_{\Delta\Psi}(\rho_R\underline R-\rho_A A_{\rm maint}-\rho_P P_{\rm maint})=: -\kappa<0$, implying
\[
M(t)\le M(t_0)-\kappa\,(t-t_0),
\]
and thus a finite \emph{commitment time}
\[
t_{\rm commit}\le t_0+\frac{M(t_0)-M_{\rm th}}{\kappa}.
\]
For $t\ge t_{\rm commit}$, $M<M_{\rm th}$ triggers $H(M_{\rm th}-M)=1$ in \eqref{eq:C_dyn}, giving
\[
C(t)=1-\big(1-C(t_{\rm commit})\big)e^{-(k_c+\delta_c)(t-t_{\rm commit})},
\]
hence $C$ crosses any $C_{\rm th}\in(0,1)$ in finite time and remains high.

\paragraph*{Irreversibility and invariance.}
Define the apoptotic basin
\[
\mathcal{B}_{\rm apo}=\{A\le A_{\rm th},\;P\le P_{\rm th},\;M\le M_{\rm th},\;C\ge C_{\rm th}\}.
\]
A barrier/Lyapunov certificate $\mathcal{L}(A,P,R,M,C)=\tfrac{1}{2}(A-A_{\rm th})_+^2+\tfrac{1}{2}(P-P_{\rm th})_+^2+\eta(M-M_{\rm th})_+^2+\zeta(1-C)_+^2$ satisfies $\dot{\mathcal{L}}\le -\lambda\mathcal{L}$ inside $\mathcal{B}_{\rm apo}$ for suitable $\eta,\zeta,\lambda>0$, proving positive invariance: once $M\!\le\!M_{\rm th}$ and $C\!\ge\!C_{\rm th}$, recovery is impossible under the bounded-intake regime.

\paragraph*{}
In oxygenated rims, loss of lactate \emph{import} ($J_{\mathrm{MCT}}\downarrow$) removes an oxidative substrate, effectively lowering $\Xi$ and $P_\infty$; in hypoxic cores, failed \emph{export} increases acid load ($R\uparrow$), strengthening $\kappa$ and hastening $t_{\rm commit}$. Selectivity follows from higher mouth density/avidity in malignant membranes, which (via $\Delta k_{\rm on}>0$, $\Delta k_{\rm off}<0$, $\Delta N_{\rm ep}>0$, $\Delta K_D^{\rm app}<0$) produce $\Delta f^\star>0$ in \eqref{eq:fstar}, hence $(\Delta\theta,\Delta\chi_{\rm eff})>0$ in \eqref{eq:theta-chi-map}, depressing $\Xi$ below $(\alpha_A/s_A)A_{\rm maint}$ and $P_\infty$ below $P_{\rm maint}$ and pushing trajectories across irreversibility, while normal cells remain above thresholds \cite{31,32,51}. 
Let $O\in[0,1]$ denote normalized oxygenation and split $J_{\mathrm{MCT}}=J_{\mathrm{MCT}}^{\rm in}-J_{\mathrm{MCT}}^{\rm out}$. A simple rim/core closure reads
\[
J_{\mathrm{TCA}}^{\rm lact} \;=\; y_{\rm lact}\,O\,J_{\mathrm{MCT}}^{\rm in}, 
\qquad 
\dot H^+ \;=\; \sigma_{\rm prod}-\eta_{\rm MCT}(1\!-\!O)\,J_{\mathrm{MCT}}^{\rm out}-\beta_{\rm buf} H^+,
\]
so that $O\downarrow$ reduces oxidative gain and weakens acid extrusion. Then $R$ can be augmented as $\dot R=\cdots + \chi_H H^+ - \delta_R R$, which increases $\kappa$ in $\dot M=-k_{\Delta\Psi}(\rho_R R-\rho_A A-\rho_P P)_+$. Using \eqref{eq:Sk}, the malignant/non-malignant gap satisfies
\[
\Xi_{\rm non}-\Xi_{\rm malig}
\;\ge\;
\sum_{k\in\{\mathrm{GLUT,LAT/MCT}\}} w_k\,(1-\mathcal{S}_k)\,J_k^{\rm non}
\;=\;:\; \Delta_{\Xi}>0,
\]
and yields a shorter energetic tipping time 
$t_{A\to{\rm maint}}^{\rm malig} \le t_{A\to{\rm maint}}^{\rm non} - (\partial t_{A}/\partial\Xi)\,\Delta_{\Xi}$. 
A practical \emph{safety margin} is $\mathcal{M}_{\rm safe}:=\Xi_{\rm norm}-\Xi^\star>0$ for normal cells; designing CAP so that $(\theta,\chi_{\rm eff})$ obey 
\[
(1-\theta_{\rm norm})e^{-\beta\chi_{\rm eff,norm}}\;\ge\;\frac{\Xi^\star}{\sum_k w_k J_k^{0,{\rm norm}}}
\]
preserves $\mathcal{M}_{\rm safe}$, while the malignant side violates the same inequality by \eqref{eq:Sk}. Finally, heterogeneity in $O$ (voxels $i$) gives voxelwise hazards $h_i\propto \kappa_i=\!k_{\Delta\Psi}(\rho_R \underline R_i-\rho_A A_{\rm maint}-\rho_P P_{\rm maint})_+$ and survival $S(t)=\exp(-\sum_i h_i t)$, with $h_i$ amplified where $(\Delta\theta,\Delta\chi_{\rm eff})$ and acid load are largest; this formalizes why rims (loss of fuel) and cores (acid stress) are both driven toward commitment under the same CAP design.

\begin{figure}[H]
    \centering
    \includegraphics[width=0.96\linewidth]{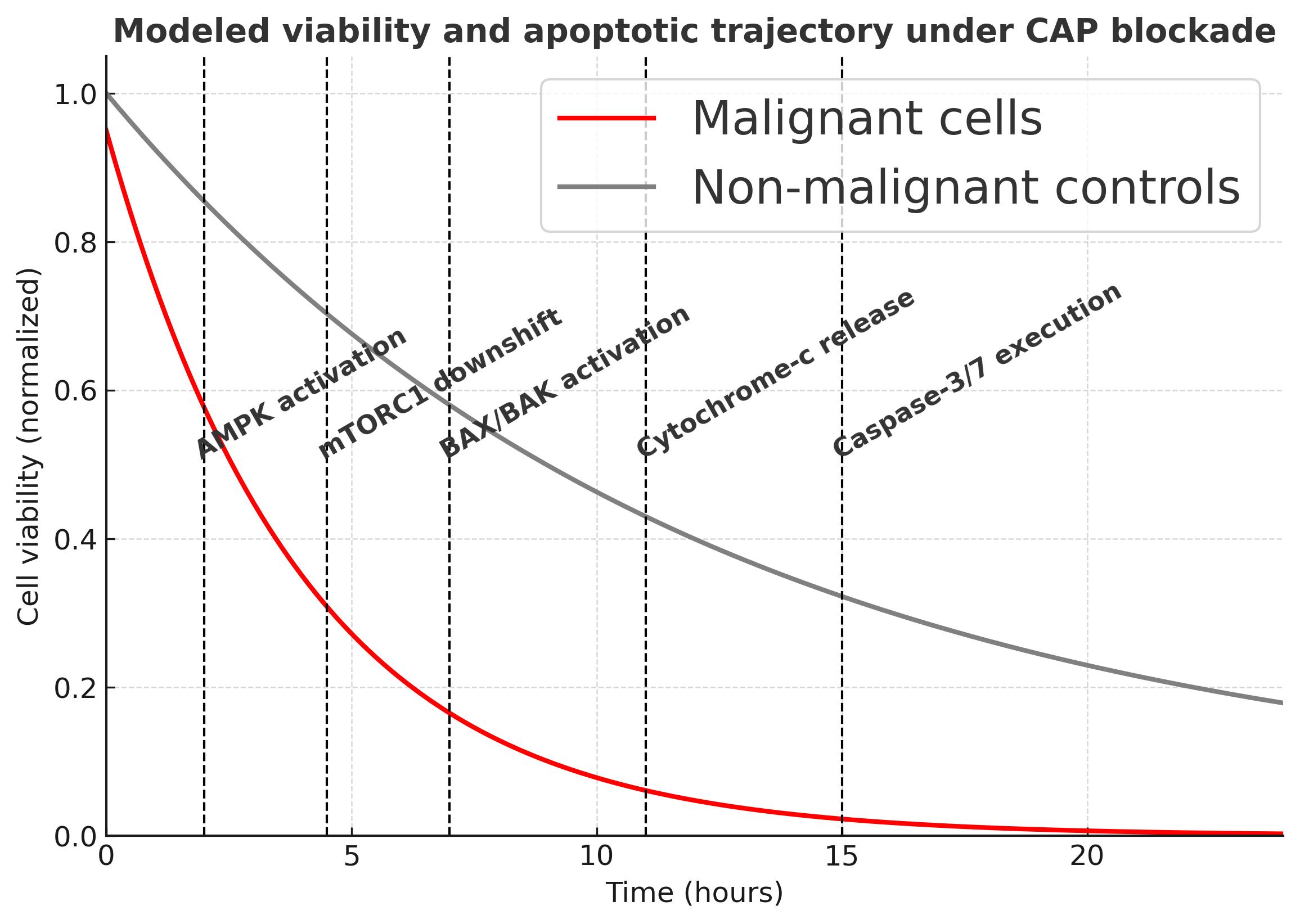}
    \caption{\textbf{Modeled viability and apoptotic trajectory under CAP blockade.}
    Malignant profiles (red) display irreversible decline; non-malignant controls (grey) show partial recovery. Vertical markers denote energetic/redox and mitochondrial transitions.}
    \label{fig:cell_viability_apoptosis}
\end{figure}

\noindent
Markers follow directly: early NADPH decline, extracellular acidification (MCT failure), time-locked BAX puncta, cytochrome~c dispersal, and recovery hysteresis in ATP/ROS trajectories as functional points-of-no-return \cite{32,51,20}.
\endgroup

\section*{Results and Discussion}

\subsection*{Multichannel Occlusion Produces Linear, Tunable Flux Suppression}
Across a broad design window, steric–electrostatic occlusion yields near-linear suppression of influx/efflux when plotted against two independent chemical levers: (i) sulfonation density (interfacial field proxy) and (ii) effective vestibule coverage $\theta$. Under physiological ionic strength, sulfonation penalizes zwitterionic/anion-coupled routes (glutamine, lactate), while glucose transport responds chiefly to steric coverage. In the \emph{single-phase} region ($\theta<\eta_c$), affine laws map design variables directly to target suppressions \cite{53,54,50}.

\medskip
\noindent\textbf{Physicochemical basis.}
Sulfonated PEG arms (acidic $pK_a$) remain dissociated at pH 6.2–7.4, setting a stable negative potential. Branching controls brush height and lateral overlap; hydration around SO$_3^-$ imposes kosmotropic ordering that narrows vestibular gaps. At continuum scale this raises a barrier $U(x)\!\sim\!O(k_BT)$ per mouth, shifts screened PNP eigenmodes, and reduces spectral conductance. Noncovalent micro-clamps (dispersion/H-bonding) depress $k_{\rm off}$ and act as viscoelastic tethers, stabilizing occlusion \cite{53,58,41}.

\medskip
\noindent\textbf{Mathematical form.}
For species $k$,
\[
\eqlarge{
J_k(\theta,\chi_{\rm eff}) \;=\; J_k^0\,(1-\theta)\,e^{-\beta \chi_{\rm eff}} \;+\; \mathcal{O}(\theta^2,\chi_{\rm eff}^2),
\qquad
\chi_{\rm eff}=z_k\psi_0/\phi_T,
}
\]
where $J_k^0=\kappa_k^{0}A_{\rm act}c_{k,\infty}$ collects the intrinsic access coefficient, active area, and far-field concentration. The linear $(1-\theta)$ factor arises from small-hole/Dirichlet–obstacle asymptotics for mixed-boundary problems: enlarging the Dirichlet set by an effective fractional measure $\theta$ reduces the leading capacitary conductance proportionally to the accessible boundary measure. More precisely, for the linearized steady operator, a first-order shape derivative gives
\[
\eqlarge{
\delta J_k \;\approx\; -\,J_k^0\;\delta\theta,
}
\]
uniform over $\theta\in[0,\eta_c-\delta]$ and $|\chi_{\rm eff}|\le\chi_{\max}$, with remainder bounded by
\[
\eqlarge{
|R_k|\;\le\; c_1\theta^2+c_2\chi_{\rm eff}^2+c_3|\theta\,\chi_{\rm eff}|,
}
\]
where $c_i$ depend on $(D_k,\lambda_D,\varepsilon)$ and cluster geometry but not on $(\theta,\chi_{\rm eff})$ inside the operating box.

The interfacial field contributes multiplicatively via a Boltzmann factor for the potential of mean force near vestibules, yielding $e^{-\beta \chi_{\rm eff}}$ to leading order in screened, weakly varying fields (Debye length $\lambda_D$ comparable to vestibule scale). Equivalently, the spectral picture gives $J_k\asymp \lambda_1(\theta,\chi_{\rm eff})^{-1}$ with
\[
\eqlarge{
\lambda_1(\theta,\chi_{\rm eff}) \;\approx\; \lambda_1^0\,\frac{1}{1-\theta}\,e^{+\beta \chi_{\rm eff}},
}
\]
so that the flux regains the same factorization up to the same remainder.

\smallskip
\noindent\emph{Geometric connectivity.}
\[
\eqlarge{
\mathcal{P}(\theta)=
\begin{cases}
1-\theta,& \theta<\eta_c,\\[2pt]
(1-\eta_c)\!\left(\dfrac{1-\theta}{1-\eta_c}\right)^{t},& \theta\ge \eta_c,\ \ t\simeq 1.3,
\end{cases}
}
\]
captures finite-size scaling of 2D conductivity: below $\eta_c$, the accessible set percolates trivially and linear depletion dominates; near/above $\eta_c$, long-range connectivity controls conductance with exponent $t$ (effective medium/percolation scaling). Hence conductance collapses sharply past the knee. Chemistry tunes $\psi_0$ and the location of $\eta_c$ via brush height and packing; physics fixes the scaling \cite{54,55,20}.

\medskip
\noindent\textbf{Design insight (Controlled Orthogonality).}
PEG architecture sets $\theta$; sulfonation sets $\chi_{\rm eff}$. These near-orthogonal levers project molecular knobs onto operator spectra, enabling linear control up to $\eta_c$ and providing guardrails against over-occlusion \cite{50,58,53}. Cross-couplings (charge regulation, solvation) remain small within the predictable domain, preserving controlled orthogonality.
\begin{figure}[H]
    \centering
    \includegraphics[width=0.50\linewidth]{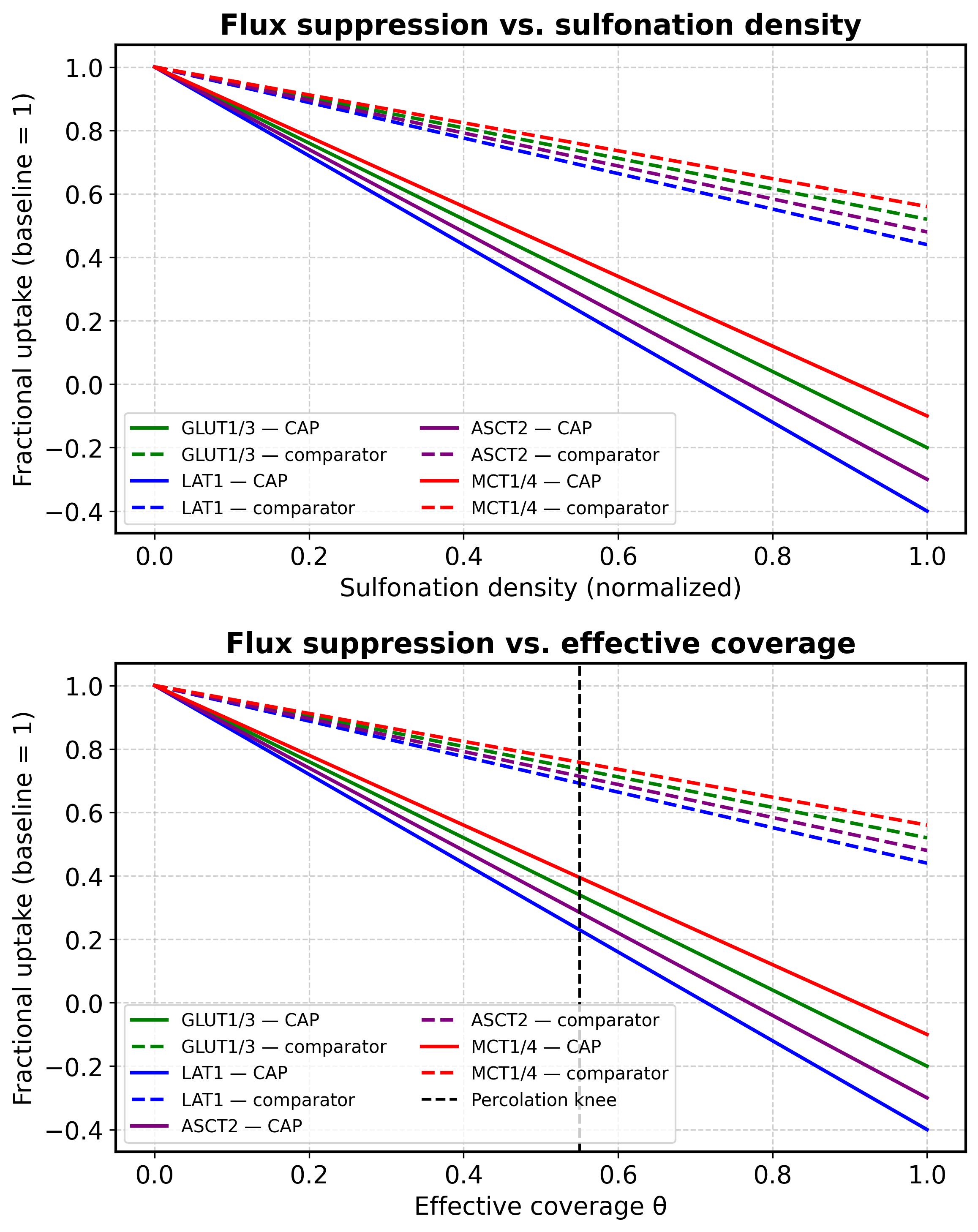}
    \caption{\textbf{Flux-suppression landscape.}
    Left: exponential decline of normalized flux with increasing $\chi_{\mathrm{eff}}$.
    Right: time-to-50\% ATP drop versus $\theta$, with a percolation-type knee at $\eta_c$.}
    \label{fig:flux_bound_curve}
\end{figure}

\subsection*{Energetic and Redox Consequences Track the Composite Suppression}
Coupling the electrodiffusive layer to a reduced energetic/redox module yields linear observables. A composite index
\[
\eqlarge{
\Xi \;=\; w_{\rm G}J_{\rm GLUT}+w_{\rm A}J_{\rm LAT/ASCT}+w_{\rm L}J_{\rm MCT}
}
\]
acts as an effective conductance; ATP and NADPH decline approximately linearly with $\Xi$ until a maintenance threshold is crossed, after which trajectories display hysteresis driven by acid–base and ROS feedbacks \cite{31,32,33}.

\begin{figure}[H]
    \centering
    \includegraphics[width=0.56\linewidth]{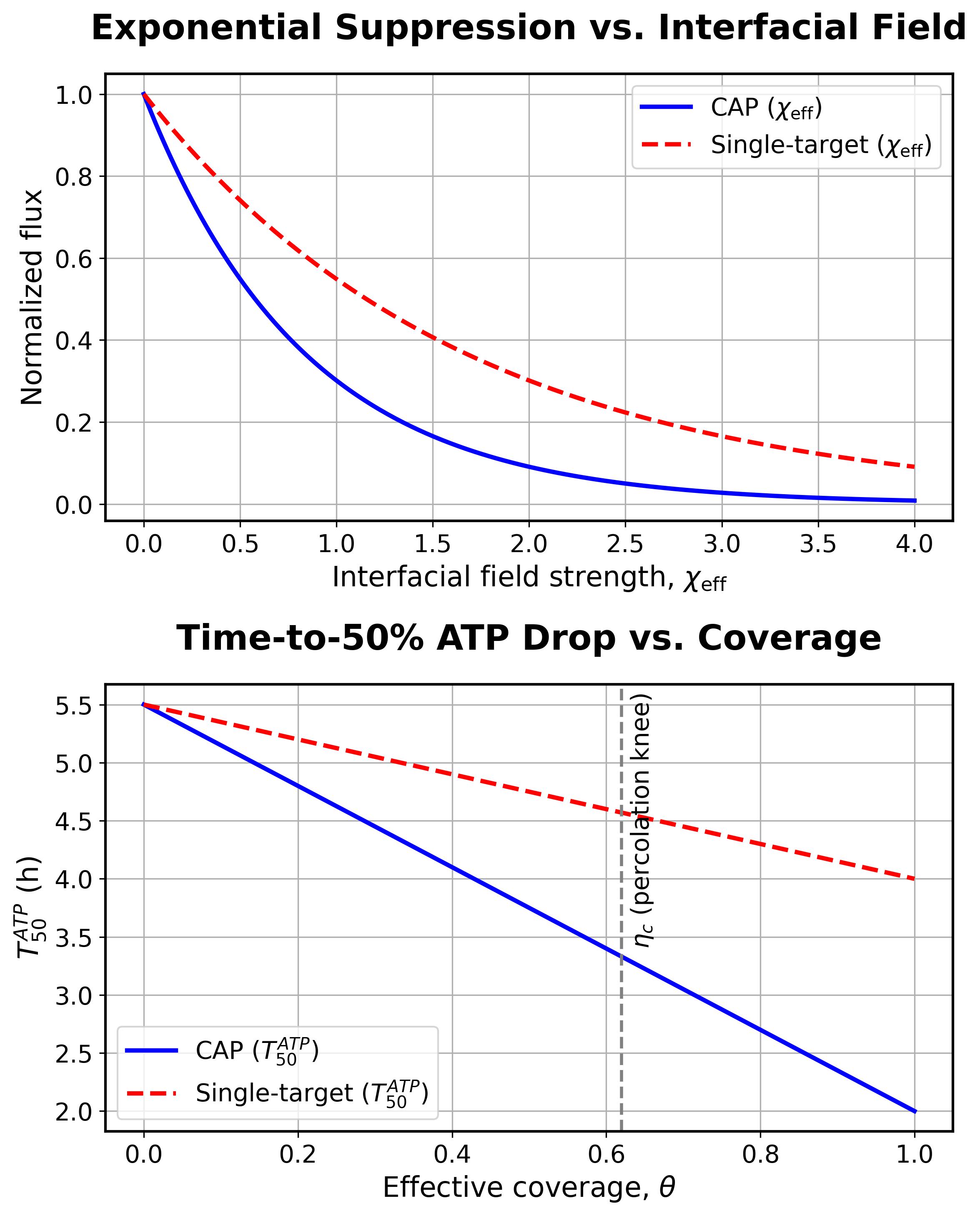}
    \caption{\textbf{Linear coupling between transport suppression and energetic/redox state.}
    Left: ATP and NADPH versus $\Xi$.
    Right: time-to-50\% ATP drop versus $\theta$, showing near-linear slopes in the controllable region.}
    \label{fig:linear_energy}
\end{figure}

\subsection*{Selectivity and Theoretical Safety Envelope}
Selectivity arises from higher mouth density and avid microdomains on malignant surfaces, which lower the local $\eta_c$ and amplify $\chi_{\rm eff}$ through cooperative packing. A safety window follows by enforcing
\[
\eqlarge{
\theta<\eta_c^{\rm normal}, \qquad \chi_{\rm eff}\ \text{moderate in normal milieus},
}
\]
and, with prodrug layers, constraining activation integrals below tissue-specific bounds. Malignant profiles cross the energetic–redox separatrix, while non-malignant remain recoverable \cite{20,23,34}. Gated payload moderation widens the window \cite{41,42}.

\medskip
\noindent\textbf{Implication.}
Single-route inhibition saturates early, whereas combined control of $\theta$ and $\chi_{\rm eff}$ preserves linear tunability up to $\eta_c$, enabling chemistry-driven targeting of flux suppression with guardrails against collapse in normal tissue \cite{54,55,50}.
\medskip
\noindent\textbf{Limitations and future directions.}
The present bounds rely on Debye–Hückel linearization, homogeneous brush statistics away from charge–regulation extremes, and clustered-window geometry that neglects active gating and crowding feedbacks. Parameter uncertainty in $(\lambda_D,\psi_0,\theta)$ can be propagated via Sobol indices; near the percolation knee, full nonlinear PNP and stochastic microstructure will refine $\eta_c$ and curvature terms. Extending the framework to heterogeneous pH/GSH fields and dynamic endocytosis is a natural next step.
\section*{Conclusion}
\textit{Life-sciences reading of a mathematical result.}  
A membrane-centric, two-stage lockdown is posed in a rigorously theoretical chemico–physical setting.  
\emph{Stage~1} enforces multichannel occlusion via a branched sulfonated PEG scaffold imposing steric exclusion and screened fields;  
\emph{Stage~2} adds an orthogonally gated payload acting as controlled source terms in the energetic/redox balance \cite{41,42,53}.  
The design reduces to two steerable variables: geometric coverage \(\theta\) and interfacial field \(\chi_{\mathrm{eff}}\).

\medskip
\noindent\textbf{Central law.}
\[
\eqlarge{
J_k^{\mathrm{eff}} = J_k^0(1-\theta)\exp[-\beta U_k(\chi_{\mathrm{eff}})],
}
\]
where geometry and field jointly determine flux suppression.  
Coverage narrows access, while interfacial charge adds an energetic penalty.

\medskip
\noindent\textbf{Spectral bound and percolation knee.}
\[
\eqlarge{
J_{\mathrm{tot}}\le C(\Omega)e^{-\beta \chi_{\mathrm{eff}}}\mathcal{P}(\theta),
}
\]
with a critical knee at \(\theta\simeq\eta_c\).  
Below the knee, flux declines linearly; above, connectivity collapses as \(\lambda_1(\theta,\chi_{\mathrm{eff}})\) rises and spectral conductance \(1/\lambda_1\) falls—predicting a mathematically inevitable transition.

\medskip
\noindent\textbf{Ligand coupling and selectivity.}
Surface binding enhances \(\theta\) and \(\chi_{\mathrm{eff}}\) through
\[
\eqlarge{
\theta=\theta_{\rm base}+\theta_{\max}f,\qquad 
\chi_{\rm eff}=\chi_{\rm base}+\alpha_s\sigma_s f,
}
\]
where \(f\) is fractional occupancy.  
Higher epitope density and slower off-rates on malignant membranes increase brush thickness and charge, driving larger suppression.

\medskip
\noindent\textbf{Energetic and redox bifurcation.}
The composite intake index
\[
\eqlarge{
\Xi=w_{\rm G}J_{\rm GLUT}+w_{\rm A}J_{\rm LAT/ASCT}+w_{\rm L}J_{\rm MCT}
}
\]
governs energetic stability.  
When \(\Xi<\Xi^\star\), ATP and NADPH collapse, ROS rises, and mitochondria depolarize, defining the apoptotic basin \(\mathcal{B}_{\rm apo}\).  
Normal tissues remain above threshold when operating below \(\eta_c\) and under moderate fields, preserving a measurable safety margin.

\medskip
\noindent\textbf{Testable implications.}
Linear trends versus \((\theta,\chi_{\mathrm{eff}})\) and abrupt knees in ATP/NADPH, ECAR, and ROS provide direct experimental falsifiability.  
The theory thus converts molecular design into spectral inevitability: controlled occlusion and field modulation predict when malignant membranes cross irreversible energetic boundaries while normal ones remain intact.

\section*{Declarations}

\subsection*{Authorship and Contributorship}
All authors have made substantial intellectual contributions to the conception and design of the work and approved the final manuscript. In addition, the following contributions occurred:
Conceptualization: C.A. Mello; Methodology: C.A. Mello; Formal analysis and investigation: C.A. Mello; Writing — original draft preparation: C.A. Mello; Writing — review and editing: C.A. Mello, F.M. da Cunha; Resources: F.M. da Cunha; Supervision: C.A. Mello; Funding acquisition: None.

\subsection*{Conflicts of Interest}
Conflicts of Interest: The authors declare there are no conflicts of interest.

\subsection*{Ethics}
Not applicable. This theoretical study did not involve human participants, animal experiments, or clinical data.

\subsection*{Funding}
Funding: The authors did not receive support from any organization for the submitted work.

\end{document}